\documentclass{article}

\usepackage{arxiv}

\usepackage[utf8]{inputenc} 
\usepackage[T1]{fontenc}    
\usepackage{hyperref}       
\usepackage{url}            
\usepackage{booktabs}       
\usepackage{amsfonts}       
\usepackage{nicefrac}       
\usepackage{microtype}      
\usepackage{lipsum}		
\usepackage{multicol}
\usepackage{graphicx}
\usepackage{natbib}
\usepackage{doi}
\usepackage{float}
\usepackage{listings}
\usepackage{hyperref}
\usepackage{svg}
\usepackage[most]{tcolorbox}

\tcbuselibrary{skins,breakable,listings}

\newenvironment{example}[1][Example]{%
    \begin{quote} 
    \textbf{#1:} \itshape
}{%
    \end{quote}
}

\title{Prompt engineering and framework:\\ implementation to increase code reliability based guideline for LLMs}

\date{2025} 					

\author{
  Rogelio Cruz$^*$  \\
  \texttt{rogelio.cruz.romero@ibm.com} \\
  \And
  Jonatan Contreras$^*$  \\
  \texttt{jcontreras@ibm.com} \\
  \And
  Francisco Guerrero$^*$\\
  \texttt{francisco.guerrero2@ibm.com}  \\
  \And
  Ezequiel Rodriguez$^*$ \\
  \texttt{ezequiel.code@ibm.com} 
  \And
  Carlos Valdez$^*$  \\
  \texttt{carlos.ignacio.valdez.aguilar@ibm.com}  \\
  \And
  Citlali Carrillo$^*$\\
  \texttt{citlali@mx1.ibm.com} \\
  \thanks{Equal Contribution. IBM Technologies, Guadalajara, México}
}

\renewcommand{\headeright}{}

\hypersetup{
pdftitle={A template for the arxiv style},
pdfsubject={q-bio.NC, q-bio.QM},
pdfauthor={David S.~Hippocampus, Elias D.~Striatum},
}

\begin{document}

\maketitle

\begin{abstract}
In this paper, we propose a novel prompting approach aimed at enhancing the ability of Large Language Models (LLMs) to generate accurate Python code. Specifically, we introduce a prompt template designed to improve the quality and correctness of generated code snippets, enabling them to pass tests and produce reliable results. Through experiments conducted on two state-of-the-art LLMs using the HumanEval dataset, we demonstrate that our approach outperforms widely studied zero-shot and Chain-of-Thought (CoT) methods in terms of the Pass@k metric. Furthermore, our method achieves these improvements with significantly reduced token usage compared to the CoT approach, making it both effective and resource-efficient, thereby lowering the computational demands and improving the eco-footprint of LLM capabilities. These findings highlight the potential of tailored prompting strategies to optimize code generation performance, paving the way for broader applications in AI-driven programming tasks.
\end{abstract}

\section{Introduction}
\label{sec:intro}
Code generation has surged as a groundbreaking application of Large Language Models (LLMs); the large pre-trained models, such as, OpenAI Codex \citep{chen2021evaluatinglargelanguagemodels}, Code LLAMA \cite{codellamaopenfoundation} or IBM's Granite \cite{mishra2024granite}, AlphaCode \citep{AlphaCode} or  InCoder \citep{fried2023incodergenerativemodelcode}, serve as benchmarks for generating functioning programs from natural language descriptions. The ability of these models to generate code, descriptions and translations, have lean on their extensive training over their vast code datasets \citep{faines2024Huan,lu-etal-2022-fantastically}. Unlike traditional natural language generation tasks, code generation requires strict adherence to syntax, logical correctness, and domain-specific constraints, presenting unique challenges and opportunities for innovation \citep{ridnik2024code}. 

Accurate prompts bridge the gap between natural language instructions and machine interpretation, ensuring optimal model performance; due to the nature and context of the task it is inevitable that the requirements written by users might be ambiguous or insufficient \cite{clarifyGPT}. For instance, AlphaCode excels in competitive programming scenarios, but its success depends heavily on prompts that clearly delineate problem constraints and expected outputs. Similarly, InCoder's pass$@$1 score of approximately 15\% \cite{fried2023incodergenerativemodelcode} highlights its reliance on in-context information, necessitating prompts that provide ample examples and partial solutions.

Some approaches to address the effectiveness of code generation, such as zero-shot \citep{radford2019language}, in-context learning \citep{in-context}, RAG \citep{RAG}, and task-specific strategies have demonstrated superior performance compared to smaller, fine-tuned models. Other approaches, like \cite{clarifyGPT, he2024codegenerationassessingcode} have leveraged LLM's performance by implementing questions or post-conditions to improve the model's understanding of the task. However, these approaches, while effective, can be costly-effective due to the number of interactions with the LLM leading to a wide amount of resource usage.

In this paper, we propose a prompting framework, that addresses the trade-offs between token efficiency and accuracy. We compare the effectiveness of our approach with Zero-Shot learning, and Chain-of-Thought (CoT) techniques, showing how instruction enhancement influence the capabilities of models like Code Llama \cite{codellamaopenfoundation} and IBM Granite \cite{mishra2024granite}.

We evaluate the effectiveness of the prompt framework in terms of overall accuracy, matching the returned results using the dataset \textsc{HumanEval}  of hand-written problems. We use the Pass@k metric; where $k$ code samples are generated per problem, and a problem is considered solved if any sample passes the unit tests, and the total fraction of problems solved is reported \citep{chen2021evaluatinglargelanguagemodels}. 

The main contributions of this paper are as follows:
\begin{itemize}
    \item We propose a novel prompting technique that surpasses the zero-shot baseline in terms of Pass@k accuracy on HumanEval. By incorporating task-specific contextual cues without resorting to verbose multi-step reasoning, our method achieves competitive results with fewer resources.
    \item Unlike chain-of-thought prompting, which increases computational overhead due to its reliance on extensive intermediate reasoning steps, our approach reduces token consumption while maintaining—or exceeding—the performance benefits
    
    \item The proposed method is adaptable across a range of LLMs, including  Code Llama and  IBM Granite. Our experimental results demonstrate consistent improvements in HumanEval Pass@k scores while preserving model scalability.

\end{itemize}

\paragraph{Paper Outline} Section \ref{sec:background} presents relevant context of LLMs and prompt optimization. Details and design of TBD are discussed in Section \ref{sec:TBD}. Experiments and results are presented in Section \ref{sec:experiments}. Our conclusions are provided in Section \ref{sec:conclusion}.

\section{Background}
\label{sec:background}
In the following section, we briefly explain the background of LLMs
for code generation and prompt engineering, then report the
most related work in the context of prompt engineering of
LLM for code.

\subsection{Code Generation and Prompt Tuning}
Transformer-based pre-trained language models are developed in two main phases: pre-training and fine-tuning. During pre-training, a transformer model is exposed to vast amounts of unlabeled data, which allows it to learn general language patterns and structures. In the fine-tuning phase, the model is further trained on labeled datasets specific to a task, enabling it to apply its understanding to more targeted outputs, such as generating syntactically correct code, adhering to specific programming conventions, or solving domain-specific challenges like debugging or algorithm design. This phase helps the model focus on nuanced tasks, such as understanding detailed function descriptions or producing optimized solutions in languages like Python, Java, or C++.

While fine-tuned models excel in task-specific performance, the process of fine-tuning comes with substantial computational and economic costs. Full fine-tuning requires updating billions of parameters, demanding extensive memory and computational resources.  To address these challenges, prompt tuning \cite{HAN2022182, lester-etal-2021-power, li-liang-2021-prefix}  has emerged as an efficient alternative, serving as a case study in achieving targeted model improvements with significantly reduced costs. 

Prompt tuning re-adapts pre-trained language models to specific tasks without modifying their internal weights. Instead of training the entire model, prompt tuning optimizes a small set of additional parameters; typically learnable embeddings, concatenated to the input prompt. It employs a prompt template $f_{prompt}(x)$ to reconstruct the original input $x$, producing new input $x'$. A good example of this technique is Prompt Tuning with Rules (PTR) \cite{HAN2022182}, where $f$ is designed as a set of structured rules or templates that are embedded into the prompt to guide the behavior of a pre-trained model. 

Broader techniques  like Zero-Shot, In-Context Learning, and Chain of Thought reasoning; further refine how final users interact with models, enabling them to perform complex tasks without task-specific fine-tuning. These methods leverage strategic input design to maximize model performance and flexibility, facilitating more effective outcomes across a variety of domains.

\subsection{Prompt frameworks}

\paragraph{Zero-Shot Learning} relies on the model's pre-trained capabilities without additional examples or explicit fine-tuning. While powerful, it often struggles with complex tasks that require domain-specific knowledge or precise outputs. For instance, a rule might specify the required output format (e.g., a Python function with a clear docstring), reducing ambiguity and enhancing zero-shot performance without examples.

\begin{example}
    In code generation, a zero-shot task could be improved by including a structured prompt:
    \begin{verbatim}
    Write a Python function to calculate the factorial of a number.
    Ensure the function has a docstring and handles invalid input gracefully.      
    \end{verbatim}
\end{example}

\paragraph{Chain of thought (CoT)} prompting explicitly encourages step-by-step reasoning by having the model break down a task into intermediate steps, improving logical reasoning and task accuracy. Rules can complement CoT by enforcing structure within the reasoning process. For example, a rule could specify that intermediate steps must be executable or logically valid in programming tasks.

\begin{example}
    Write a Python function to compute the nth Fibonacci number
    \begin{verbatim}
    Rules:
    1. Use recursion with memoization.
    2. Include a base case for n = 0 and n = 1.
    
    Step-by-step reasoning:
    1. Define a recursive function.
    2. Add memoization to improve efficiency.
    3. Handle base cases for n = 0 and n = 1.     
    \end{verbatim}
\end{example}

While CoT increases performance, it can also add to the computational cost, as more tokens are needed to express intermediate reasoning steps. This requires careful management of token limits, especially for large models with strict token constraints. CoT heavily relies on context—models must be trained or prompted effectively to generate meaningful reasoning steps. Without proper context or the right kind of structured prompt, CoT reasoning might not provide the desired results.

Other widely studied frameworks include One-Shot, Few-Shot Prompting \cite{brown2020language}, Self-Consistency \cite{wang2024prompt}, Tree-of-Thoughts \cite{yao2023tree}, Retrieval-Augmented Generation (RAG) \cite{RAG}, and Automated Prompt Engineering (APE) \cite{APE}. However, this paper focuses on Zero-Shot Learning and Chain of Thought (CoT), as these are directly applicable to code generation tasks. While the aforementioned frameworks offer potential improvements, they either introduce additional computational complexity or are less suited for code-based applications. Therefore, the discussion is centered on methods with clear relevance to programming and code synthesis.

\section{Hybrid Approach}
\label{sec:TBD}
While zero-shot prompting offers simplicity and generality, and chain-of-thought prompting excels in fostering detailed reasoning, the space between these two techniques presents an opportunity for innovation. In this section we introduce a new approach carving out a middle ground that leverages the adaptability of zero-shot methods and the structured depth of chain-of-thought strategies introducing code quality instructions into clear sections. This enables the model to perform tasks efficiently while maintaining a clear focus on the essential steps required for the task at hand. We describe the set as follows\footnote{The items only show the description of each section and not the template itself}:

\begin{itemize}
    \item Analyze: We provide the model the core of the task, describing in a minimal form that it will receive a task and that it must return a response.  
    \item Design: If it comes up with more that one answer, then it must choose the most efficient one.
    \item Implement: The form of the output must be clean and with the use of the most adequate resources.
    \item Handle: It has to be content aware, that should predict errors and overcome code limitations.
    \item Quality: It must follow code quality rules and good practices
    \item Redundancy Check: Code generation models tend to loop over the response, here we instruct to avoid it.
\end{itemize}
 
We call this template \textbf{ADIHQ} by the template's acronym. These steps greatly enhance the precision and effectiveness of prompts for LLMs. By employing a structured methodology, abstract goals are transformed into clear, actionable outputs, resulting in more tailored and impactful responses. Since this approach merely refines the initial prompt without introducing Few-Shot examples or Chain of Thought reasoning, it improves output quality while maintaining resource efficiency.

The following example shows how ADIHQ prompt can be implemented using a zero-shot base from the HumanEval dataset \cite{chen2021evaluatinglargelanguagemodels}.
\begin{figure}[H]
	\centering
    \includegraphics[scale=0.9]{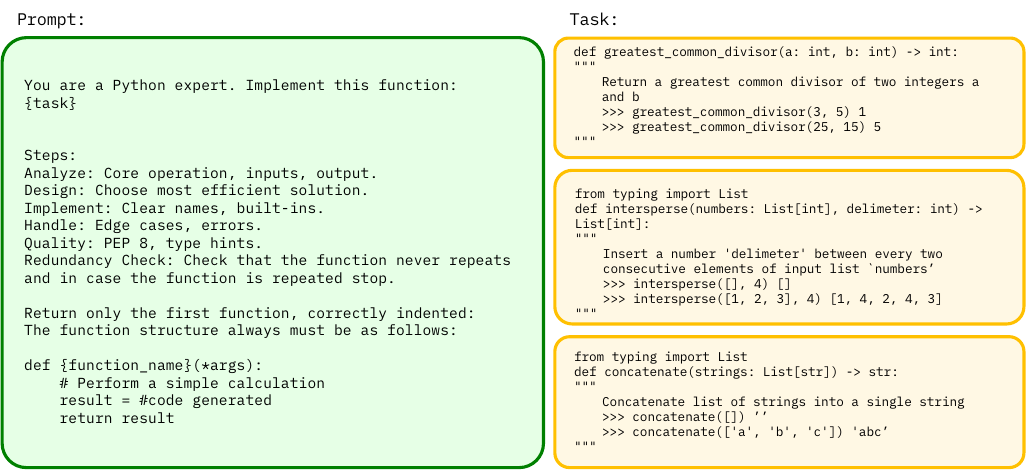}
	\caption{Sample figure caption.}
	\label{fig:fig1}
\end{figure}

 This technique is designed to balance efficiency with reasoning depth, making it particularly suited for scenarios where direct answers require nuanced understanding without the overhead of fully detailed reasoning chains. By filling this methodological gap, this approach enhances the model’s ability to generate relevant, insightful responses across a wide range of tasks.

\section{Experiments}   
\label{sec:experiments}
We conduct experiments to show the effectiveness of ADIHQ on many-class
code generation tasks and we contrast them with Zero-Shot and Chain-of-Thought frameworks. We evaluate the accuracy of the models and compare the generation time, token use, and Pass@k metric.

\subsection{Evaluation Framework}
\label{subsec:eval_framework}
In this work we evaluate functional correctness using the Pass@k metric, it measures the probability that at least one of the top $k$ generated code samples correctly solves a given programming problem.

For a programming task, a model generates multiple candidate solutions (e.g., code snippets). Pass@k checks whether at least one of these $k$ solutions meets the correctness criteria (predefined test cases). This metric accounts for scenarios where models might generate several plausible solutions, some of which could fail but others succeed.

If $n\geq k$ samples are generated per task, and $c$ is the number of correct samples that passes the unit tests, the Pass@k metric is computed as:
\begin{equation}
    Pass@k := \underset{\text{problems}}{\mathbb{E}} \left[  1- \frac{{{n-c}\choose k }} { {n\choose k}}   \right]
\end{equation}

This formula assumes that correctness is distributed uniformly among the samples, making it a more reliable measure when exact correctness probabilities are not available.

We evaluate functional correctness on the HumanEval dataset; it is a benchmark for evaluating code generation models, comprising 164 Python programming problems. Each problem includes a natural language task description, a function signature, and a suite of unit tests to assess correctness. It is designed to measure a model's functional accuracy, generalization ability, and performance in scenarios such as zero-shot and few-shot learning, which is well suitable to our approach. The dataset can be found at \url{https://www.github.com/openai/human-eval}.

\subsection{Results}
\label{subsection:results}
We conducted a qualitative analysis of the code generated by three distinct models, focusing on both the diversity and correctness of the outputs. During testing, it was observed that in some cases, the model-generated responses included multiple functions, often repeated with slight variations. To address this issue, a repetition penalty of 1.2 was applied, which helped reduce the frequency of redundant outputs. Additionally, we performed a systematic cleanup of the generated code, removing extraneous text and retaining only the first generated function for further evaluation. Additional parameters such as $temperature = 0.4$ and $top_p=0.3$ were implemented.

\begin{figure}[H]
	\centering
    \includegraphics[scale=0.8]{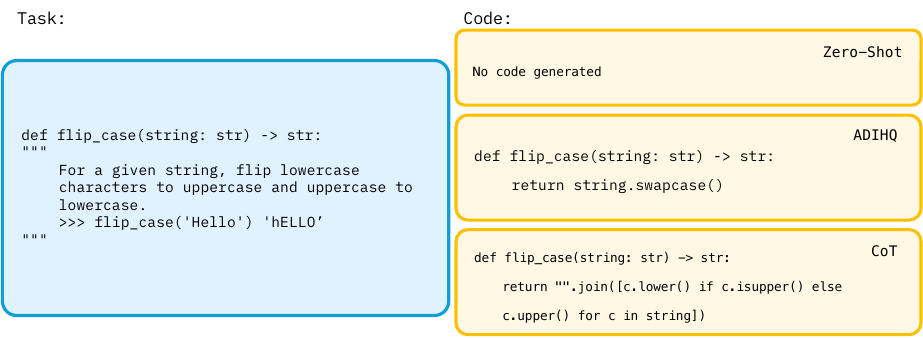}
	\caption{Code generated for flip the case of a string.}
	\label{fig:flip_case}
\end{figure}

\begin{figure}[H]
	\centering
    \includegraphics[scale=0.8]{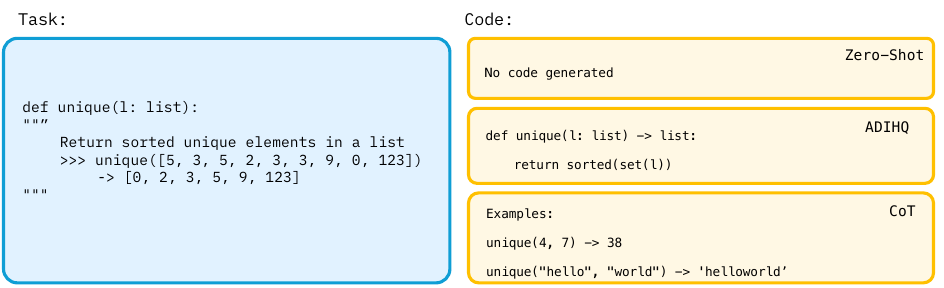}
	\caption{Code generated for sorted unique elements in a list.}
	\label{fig:unique}
\end{figure}

\begin{figure}[H]
	\centering
    \includegraphics[scale=0.8]{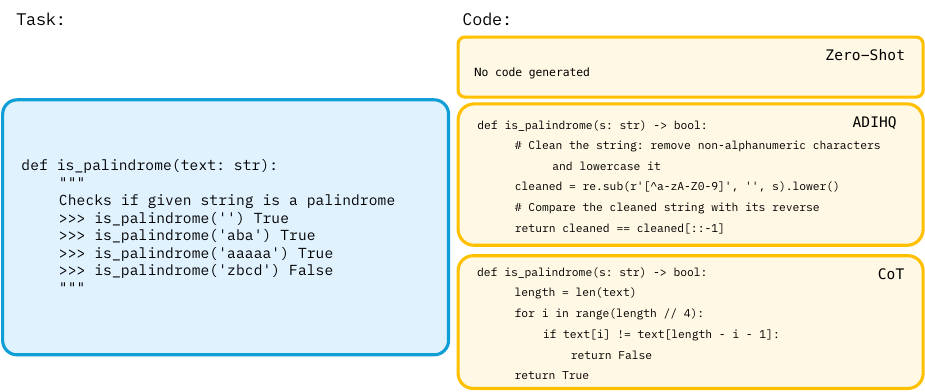}
	\caption{Code generated for returning if string is palindrome.}
	\label{fig:is_palindrome}
\end{figure}

Figures \ref{fig:flip_case}, \ref{fig:unique} and \ref{fig:is_palindrome} illustrate examples of code generated using the three approaches applied to the Granite model. Supplementary results for both the Granite and LLAMA models are provided in Appendix \ref{appendix:llama}.

For instance, in Figure \ref{fig:flip_case}, we present the generated code for \texttt{flip\_case}; involved converting a string by changing lowercase characters to uppercase and vice versa. The results show that both ADIHQ and CoT approaches successfully generated functional code. Although the same underlying model was used, the generated code differed in programming logic. Noting that both approaches passed all test cases, demonstrating their ability to generate correct solutions for this task.

The second case study focuses on the unique task (Figure \ref{fig:unique}), where the objective was to extract and sort the unique elements from a given list. In this instance, only ADIHQ produced a correct and consistent solution. The other approaches generated code that failed to meet the task requirements, resulting in incorrect outputs.

 In the third case study, the \texttt{is\_palindrome} task shown in Figure \ref{fig:is_palindrome}, required determining whether a given string is a palindrome. While the CoT approach generated code that passed some of the test cases, it failed to handle all edge cases, rendering the solution incomplete. By contrast, ADIHQ produced code that successfully passed all test cases, showcasing its reliability and correctness for this task.

In Table \ref{tab:performance_metrics} we conducted a quantitative analysis of the performance of the three approaches for two different models, Granite and LLAMA Code. This evaluation focuses on two key metrics; token usage, which measures the number of tokens generated by each approach, and Pass@k described in \ref{subsec:eval_framework} for $k=1,100$. We implemented this tests using the watsonx.ai platform, an advanced AI and data platform designed to facilitate the development, deployment, and evaluation of AI models with robust tools for tracking metrics and optimizing performance.

ADIHQ aims to improve precision in token utilization, ensuring that each token contributes effectively to generating successful solutions. To assess this, we include the column Pass$_{100}$@token in Table \ref{tab:performance_metrics}, which represents the ratio of Pass@100 to the number of tokens used. To enable meaningful comparisons, we further normalize this ratio by dividing it by the reference case, in which we assume that the tokens spent in a Zero-Shot setting result in a 100\% Pass@100 accuracy. This normalization accounts for differences in token consumption across models, providing a clearer perspective on their relative efficiency.

\begin{table}[H]
	\centering
    \caption{Experimental results of Pass@k and token usage for the three approaches in Granite and LLAMA Code models.}
	\begin{tabular}{lllll}
		\toprule
		\multicolumn{2}{c}{Zero-Shot}                   \\
		\cmidrule(r){1-5}
		Model     &  Tokens & Pass@1 & Pass@100 & Pass$_{100}$@token\\
		\midrule
        
		Granite &  234.8  & 0.05 & 0.1 & 0.1 \\
		LLAMA Code     & 259.99 & 0 & 0 & 0\\
		
		\bottomrule
        \addlinespace[3pt]
        \multicolumn{2}{c}{Chain-of-Thought}                   \\
		\cmidrule(r){1-5}
		Model         & Tokens & Pass@1 & Pass@100& Pass$_{100}$@token\\
		\midrule
		Granite & 326.58  & 0.25 & 0.3 & 0.22\\
		LLAMA Code      & 352.99 & 0.43 & 0.5 & 0.36 \\
		
		\bottomrule
        \addlinespace[3pt]
        \multicolumn{2}{c}{ADIHQ}                   \\
		\cmidrule(r){1-5}
		Model     &  Tokens & Pass@1 & Pass@100& Pass$_{100}$@token\\
		\midrule
		Granite &237.58  & 0.41  & 0.433 & 0.43\\
		LLAMA Code     &  259.99 & 0.41  & 0.4666 & 0.44\\
		
		\bottomrule
	\end{tabular}
    
	\label{tab:performance_metrics}
\end{table}
The results provide a detailed comparison of the three approaches—zero-shot, CoT, and ADIHQ—based on the Pass@k metrics and token usage. The zero-shot approach, despite leveraging code-specific models, demonstrated notably lower performance compared to the CoT and ADIHQ methodologies. Its inability to consistently generate accurate solutions highlights the limitations of relying solely on direct model outputs without additional reasoning or guidance.

In contrast, the hybrid approach consistently outperformed the CoT method, achieving both higher and more stable accuracy across the Pass@1 and Pass@100 metrics. Notably, this improvement was achieved without a significant increase in token usage. ADIHQ maintained a token consumption profile comparable to that of the zero-shot method, indicating that its enhanced reasoning and accuracy did not come at the cost of excessive computational overhead.

In contrast, when examining the Pass@k metrics, we observe that CoT achieves better performance for LLAMA Code models. However, while ADIHQ maintains consistently high precision, it does so with more efficient resource utilization. This efficiency becomes particularly evident when analyzing the Pass$_{100}$@token metric, where ADIHQ outperforms CoT, demonstrating superior token efficiency while maintaining competitive accuracy.

This balance between accuracy and efficiency underscores the effectiveness of the hybrid methodology. Minimizing token usage in LLM-based code generation has significant implications for reducing the eco-footprint of AI technologies. By cutting down computational resources required for model inference, our proposal not only enhances efficiency but also reduces energy consumption and associated carbon emissions. This aligns with broader sustainability goals, making AI-driven solutions more environmentally responsible while maintaining high-quality outputs. Furthermore, it demonstrates a scalable approach to optimizing LLM capabilities without compromising on performance or accuracy.

\section{Related Work}

Advances in large language models (LLMs) have spurred significant progress in automated code generation, with studies focusing on improving accuracy, efficiency, and adaptability. Few-shot learning has become a cornerstone in prompt optimization, where frameworks like \cite{CUI2025101762} enhance meta-learning by incorporating external knowledge to create more effective prompts. Similarly, document-driven approaches such as DocCGen \cite{doccgen} leverage structured inputs to refine prompt control and improve Python and multi-language code generation. These methods highlight the growing role of prompt design in tailoring outputs to meet diverse requirements while addressing scalability challenges.

Frameworks like FRANC \cite{siddiq2024franc} take a holistic approach to improving code generation quality. By combining static filtering for compilability, a quality-aware ranker for snippet prioritization, and targeted prompt engineering, FRANC demonstrates significant enhancements in output quality for Python and Java models with minimal computational overhead. Similarly, EPiC \cite{taherkhani2024epic} employs evolutionary algorithms to refine initial prompts, steering LLMs toward generating high-quality code with reduced interaction costs.

Security-focused approaches such as PromSec \cite{PromSec} use generative adversarial networks (gGANs) and contrastive learning to iteratively refine prompts and code, reducing vulnerabilities while enabling transferability across languages and models. ClarifyGPT \cite{clarifyGPT}, on the other hand, combines ambiguity detection with interactive refinement to improve user-driven code generation processes. While these methods demonstrate notable improvements in security and user interaction, their reliance on iterative feedback mechanisms can be resource-intensive, particularly for large-scale or complex tasks.

Other studies explore optimization and workflow orchestration to enhance LLM capabilities. For instance, WorkflowLLM \cite{fan2024workflowllm} optimizes Python function generation through prompt-tuned models, while ClarifyGPT selectively engages with users to resolve ambiguities. These frameworks complement foundational techniques such as trace-based optimization (e.g. \cite{cheng2024traceautodiff}), which uses meta-prompts to refine workflows for Python code generation. Collectively, these studies underline the critical role of prompt engineering in advancing code generation, paving the way for proposals like ours that integrate quality, security, and efficiency into a unified approach.

\section{Conclusion and Future Work}
\label{sec:conclusion}
In this paper, we introduced a novel prompt framework, ADIHQ, designed to enhance the reliability and accuracy of code generated by large language models (LLMs). Through comprehensive analysis, our study demonstrates that the proposed framework significantly outperforms existing Zero-shot and Chain-of-Thought (CoT) approaches, achieving higher accuracy in code generation while maintaining efficient resource usage. By striking a balance between accuracy and computational cost, our framework provides a scalable and effective solution for improving code generation tasks.

We summarized key findings that highlight the advantages of this approach and discussed its broader implications for researchers and practitioners. Our work provides a foundation for further exploration of prompt engineering strategies in code intelligence tasks, offering new opportunities to optimize the performance of LLMs in practical applications.

We chose to implement our approach on LLAMA and Granite due to their lower computational profiles, making them ideal for testing efficiency-focused enhancements. Unlike extensive LLMs like GPT, which demand significant computational resources and energy, these smaller models allow us to demonstrate the practicality of our method in resource-constrained environments. By optimizing token usage and enhancing performance on lightweight models, we prove that high-quality code generation can be achieved without the environmental and operational overhead of larger LLMs. This also highlights the scalability of our approach, enabling broader accessibility and integration into real-world applications where cost and energy efficiency are paramount.

In the future, we plan to extend this work by evaluating the accuracy and efficiency of our approach across a wider range of models and diverse datasets. This will allow us to assess the generalizability of the framework and identify potential enhancements for broader use cases in code generation and other domains.

\bibliographystyle{unsrtnat}
\bibliography{references.bib}  






\appendix

\section{Results LLAMA Code}\label{appendix:llama}
In Section \ref{subsection:results}, we presented the qualitative results of code generation using the Granite model. In this section, we extend the analysis to the LLAMA Code model, providing a comparative evaluation of its performance under the proposed method. These experiments are designed to assess the model's ability to address edge cases effectively, showcasing its robustness and reliability in handling diverse programming challenges. Additionally, Table 2 provides a detailed comparison of token usage across the three experiments, highlighting the computational efficiency of each approach.

\begin{figure}[H]
	\centering
    \includegraphics[scale=0.8]{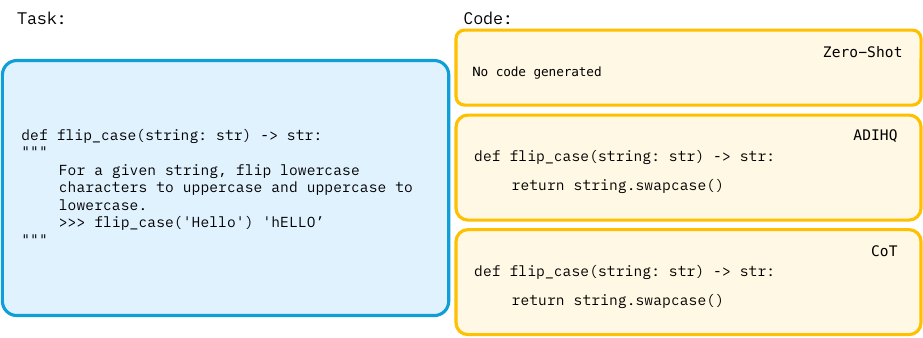}
	\caption{Code generated for flip the case of a string.}
	\label{fig:flip_case_ll}
\end{figure}
\begin{figure}[H]
	\centering
    \includegraphics[scale=0.8]{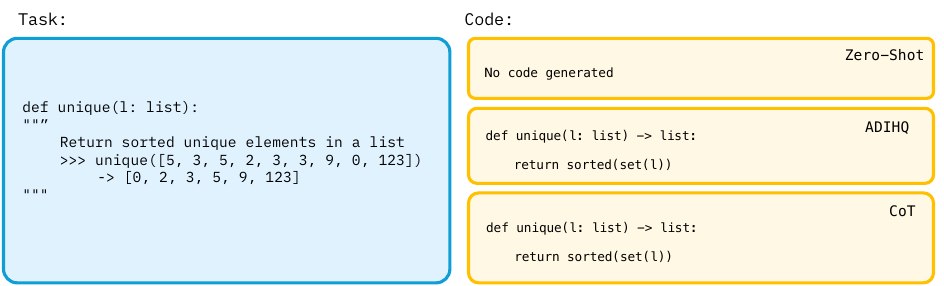}
	\caption{Code generated for sorted unique elements in a list.}
	\label{fig:unique_ll}
\end{figure}

\begin{figure}[H]
	\centering
    \includegraphics[scale=0.8]{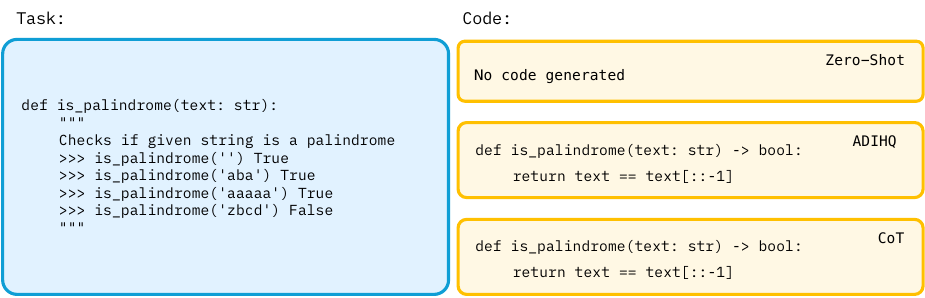}
	\caption{Code generated for returning if string is palindrome.}
	\label{fig:flip_caseis_palindrome_ll}
\end{figure}
\begin{table}[H]
	\centering
    \caption{Token spent usage for LLAMA Code}
	\begin{tabular}{lccc}
		\toprule
		\textbf{Task} & \textbf{Zero-Shot} & \textbf{ADIHQ} & \textbf{CoT} \\
		\midrule
		\texttt{flip\_case}       & 184 & 192 & 277 \\
		\texttt{unique}           & 196 & 204 & 289 \\
		\texttt{is\_palindrome}   & 216 & 224 & 309 \\
		\bottomrule
	\end{tabular}
	\label{tab:performance_metrics}
\end{table}
In these experiments, we observed that both ADIHQ and CoT approaches generated identical code that successfully handled all test cases with complete correctness. However, a notable distinction lies in their computational efficiency. As shown in Table 2, the token usage for ADIHQ is significantly lower compared to CoT, highlighting its superior resource efficiency while maintaining the same level of accuracy. This finding underscores the advantages of ADIHQ in optimizing performance without compromising correctness.

\end{document}